\documentclass[a4paper,10pt]{article}

\usepackage{graphicx}
\usepackage{verbatim}
\usepackage{textcomp}

\title{Asymmetric statistics of order books: The role of discreteness and evidence for strategic order placement}
\author{A. Zaccaria, M. Cristelli, V. Alfi, F. Ciulla, L. Pietronero}

\begin{document}

\maketitle

\begin{abstract}
We show that
the statistics of spreads in real order books is characterized by 
an intrinsic asymmetry due to discreteness effects for even or odd values of the spread. 
An analysis of data from the NYSE order book points out that traders' strategies
contribute to this asymmetry.
We also investigate this phenomenon in the framework of a microscopic model
and, by introducing a non-uniform deposition mechanism for limit orders, we
are able to quantitatively reproduce the asymmetry found in the
experimental data.
Simulations of our model also show a realistic dynamics
with a sort of intermittent behavior characterized
by long periods in which the order book is compact and liquid
interrupted by volatile configurations. The order placement strategies produce a 
non-trivial behavior of the spread relaxation dynamics which is similar to the one observed in real markets.
\end{abstract}

\section{Introduction}
\label{sec:1}

The order book is the double auction mechanism~\cite{has,harris,bouchaud} which permits
to process and store the orders placed by investors in a modern financial market.
This system is the elementary mechanism of price formation
as a consequence of the arrival of  proposals (orders) of buying or selling.
There are two classes of orders: market orders and limit
orders. The market ones correspond to orders to buy/sell at the
best available price (called best bid/ask), hence they are immediately executed.
The limit orders instead are orders to buy or sell at a given price which can be 
not necessarily the best one.
By consequence limit orders may not immediately fulfilled and then they are
stored in the order book.
The difference between the best ask and the best bid is 
defined as the spread $s$. The order prices are not continuous but discrete and
expressed in units of ticks.
Also the volume of an order is an integer multiple of a certain amount of shares.
The mid-price between best ask and best bid can be considered a conventional definition of the price of a stock. 
The majority of the spread variations are due to a limit order which falls inside the spread or 
to a market order which matches all the orders placed at the best price.
\\
In recent years the complex dynamics of the order book has attracted the scientific community attention.
One of the reasons of this increasing interest is the availability of a large amount of experimental data which has permitted an 
extensive statistical analysis of the order book properties and has revealed a number of interesting features and regularities.
The main empirical evidences of the order book are: the fat-tailed distribution of the price of 
new limit orders \cite{bouchaudI,bouchaudII,farmer_emp};
the non-trivial power law correlation of the transaction price signs~\cite{bouchaud-2008,bouchaud-signI,bouchaud-signII,farmer-longI,farmer-longII};
the peculiar shape of the limit order volume distribution with fat tails and peaked away from the best quote~\cite{bouchaudI,bouchaudII};
the non linear response of the order book to order arrivals~\cite{farmer-nature,stanley}.
\\
Moreover the microscopic market level of the order book shows the stylized facts present in financial 
markets at the aggregate level~\cite{bouchaud,mantegna,mandelbrot,cont}.
To interpret these empirical evidences a series of theoretical models has been introduced
~\cite{slanina,genoa,maslov,challet2,farmer_emp,farmer-prl}.
In particular, the models by Farmer et al. ~\cite{farmer_emp,farmer-prl}
show as even a ``zero intelligence'' mechanism can reproduce many of
the experimental features of real order books.
\\
In a previous work we have introduced a model~\cite{ob_pil} where the zero intelligence paradigm is adopted
and in this paper we are going to interpret some experimental results in the framework of this model.
\\
One of the most challenging issue in this field is the identification of new empirical features
in order to discriminate and validate the various models proposed.
In this paper we investigate the effects induced by the discrete nature of order books.
In fact, as we have already noticed, the price of the orders is not a continuous variable but it can only be a multiple of a quantity called tick which is a fraction of the currency used in the the market considered.
A first consequence of this aspect is the spontaneous emergence of asymmetries in the system.
For instance, two configurations of the order book with an even spread or an odd one (in units of ticks) are not 
{\it {a priori}} equivalent for the mechanism of deposition of limit orders inside the spread.
We investigate the fraction of odd spreads for a data set from the NYSE market
finding indeed a strong evidence for this asymmetry. However, as we are going to see in detail in the paper,
the problem is more subtle than expected and also the agent strategies play an important role to 
explain quantitatively this phenomenon. The strategic order placement is also the origin of the non-trivial
relaxation pattern observed when a spread fluctuation takes place.\\
The paper is organized in the following way.
\\
In Sec.~\ref{sec:2} we show the experimental evidence of the asymmetry of spreads.
\\
In Sec.~\ref{sec:3} we propose an interpretation of this empirical feature in terms of
a microscopic model we introduced in~\cite{ob_pil}.
\\
In Sec.~\ref{sec:4} we perform a more detailed data analysis in order to investigate further aspect
which contributes to the asymmetry.
\\
In Sec.~\ref{sec:5} our model is properly modified to take into account the new results found 
in the data analysis. 
\\
In Sec.~\ref{sec:6} a detailed investigation of the role of the agent strategies for the placement of order inside the spread is performed in the framework of our model.
\\
In Sec.~\ref{sec:7} we study the effects of strategic order placements on the spread relaxation dynamics with respect to a pure zero-intelligence mechanism.
\\
Finally conclusions and perspectives are discussed in Sec.~\ref{sec:con}.

\section{Empirical evidences}
\label{sec:2}

In this section we show the empirical evidences that have led us to
investigate the asymmetric nature of the order book.
\\
In our analysis of the order book we have considered a data set which spans a 
period of nearly 80 trading days between October 2004 and February 2005.
This data set includes a series of high-frequency (tick-by-tick) information for 20 stocks
from the NYSE market.
These stocks have been chosen to be heterogeneous in their level of 
capitalization \footnote{Market capitalization is defined as the number of shares of a company multiplied by their price. It is the simplest measure of a company's size.}. The information we have for each stock is the whole 
list of transactions and quotes, which are the prices of effective
deals and of orders respectively. Our data set lacks the information about
the whole order book apart of the best bid and the best ask. Therefore
a new quote appears only if the best bid
or the best ask have been updated.
From this data set we have reconstructed the sequence of market and limit orders
by looking to the spread variations. If the spread has increased with respect to its 
previous value we refer to the event as a market order. On the contrary, 
the event is defined as a limit order if the spread has decreased.
We assume that 
the probability that a cancellation of a limit order may change the spread is so small
to be neglected.
We have also expressed all the prices in units of ticks, so the spread
results to be an integer number.
At the end of this refinement, our data set is composed only by the series of the events in which 
the spread has changed, each labelled to be a market or a limit order.
In this paper we restrict our analysis to limit order events. 
The complete analysis, including also market orders, will be matter
of future researches.
To investigate the intrinsic asymmetry that a discrete and finite
spread generates, we have firstly analyzed how the fraction of odd
spreads depends on the average value of the spread for each stock.
\begin{figure}[t]
\begin{center}
\includegraphics[clip=true,keepaspectratio,height=6.0cm,angle=0]{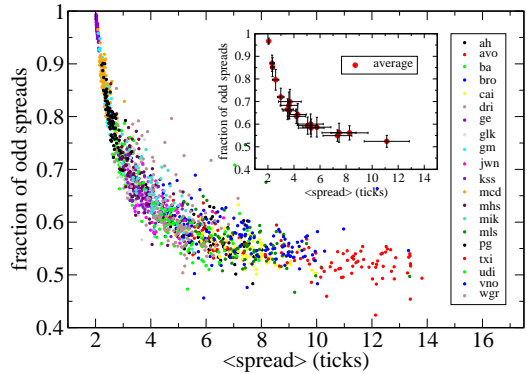}
\caption{(Color online) Fraction of odd-valued spreads (in units of ticks) vs daily average spread for different stocks. 
 We observe a systematic deviation from the symmetric case in which the fraction is $0.5$. In the inset we plot
a further average over a period of 80 days.}
\label{fig:1.1}
\end{center}
\end{figure}
In Fig.~\ref{fig:1.1} we have plotted, for each stock, the 
daily fraction of odd-valued spreads as a function of the 
average value of the spread. There are 80 different points for each stock
and also a further average over all the days is plotted in the inset.
We can observe that for almost all points 
the fraction of odd-valued spreads is larger than $0.5$
which would be the expected value if the spread was very large ($s\rightarrow\infty$).
This asymmetry is more marked for stocks with
a smaller average spread and it goes diminishing while the average
spread increases.
A small average spread usually corresponds to stocks with a 
large capitalization.
In the next sections we are going to investigate these results 
in the framework of the model we have introduced in a previous 
work~\cite{ob_pil}.

\section{A model for limit order deposition: uniform case}
\label{sec:3}

In this section we propose a simple explanation of the evidences shown in Sec.~\ref{sec:2} on the basis of the order book 
model introduced in~\cite{ob_pil}.
\\ 
First of all we briefly recall the main properties of the model. At each time step an order is placed. 
This order can be a sell or a buy order with the same probability $1/2$, and a market or a limit order with probability 
$\pi=1/3$ and $1-\pi=2/3$ respectively. A limit order is placed with an uniform distribution in the interval $]b(t), b(t)+k s(t)]$ if it is a sell order and $[a(t)-k s(t), a(t)[$ if it is a buy order, where $b(t)$ and $a(t)$ are the best bid and the best ask respectively, $s(t)$ is the spread and $k>1$ is a constant. In first
approximation $k^{-1}$ is the probability for a limit order to be put inside the spread (therefore causing a price change). This dependence on the previous spread value creates a sort 
of autoregressive mechanism for the order deposition.
The tick size and the volume of the orders are both constant and fixed ($\Delta q=1$ and $\omega=1$). Finally, a cancellation process avoids the divergence of the volume stored in the order book.  \\
In this framework we can evaluate a number of quantities, for example the probability that, given a spread $s$ at time $t$, the new spread $s'\neq s$ at time $t+\Delta t$ is even or odd, where $\Delta t$ is the time to wait to have a variation of the spread. The dependence on the parameter $k$ is removed because we consider only the conditioned probability that an event occurs. In this paper we indicate with $s$ the value of the spread before
an incoming event and with $s'$ the new value of the spread consequent to the variation. 
Here we restrict our analysis only to events due to limit order arrival.
The probability to have an odd spread in the final state turns to be dependent on the parity of the spread $s$. Straightforward calculations give the probabilities
\begin{equation}
\label{eq:3.1}
\left.
\begin{array}{l}
P(e|o,s)=\frac{1}{2}\\\\
P(o|o,s)=\frac{1}{2} 
\end{array}
\right.
\left.
\begin{array}{l}
P(o|e,s)=\frac{1}{2}\frac{s}{s-1}\\\\
P(e|e,s)=\frac{1}{2}\frac{s-2}{(s-1)}.
\end{array}
\right.
\label{ashgdjs}
\end{equation}
where $o$ and $e$ denote if the the spread is odd or even respectively. One can see that $P(o|e)>P(e|e)$: given an even spread $s$ in the initial state, the next spread  $s'$ has a larger probability to be odd rather than even. As expected, both $P(o|e)$ and $P(e|e)$ tend to $1/2$ for $s \rightarrow \infty$. This simple argument gives a first explanation of the empirical evidence of an excess of odd spreads shown in the previous section.

\section{Data analysis}
\label{sec:4}

\begin{figure}[t]
\begin{center}
\includegraphics[clip=true,keepaspectratio,height=6.0cm,angle=0]{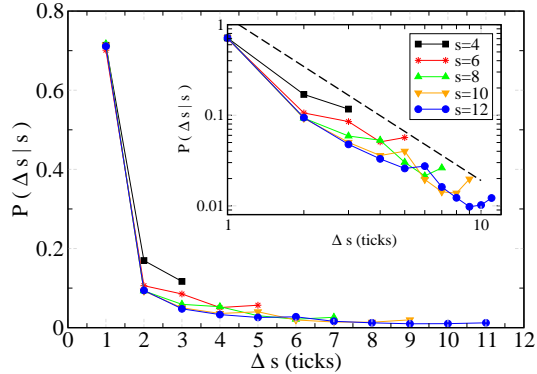}
\caption{(Color online) Probability to have a variation $\Delta s$ of the spread given an initial
$s$ for different spread values (empirical data).
The most probable variation is always $\Delta s=1$. The probabilities of the
other variations ($\Delta s=2,3,...$) are weakly decreasing functions and, in first approximation, can be considered as constant.
The dashed line is the power law with exponent $1.8$ found in \cite{ponzi} for London Stock Exhange. In both cases $\Delta s=1$ is highly preferred.}
\label{fig:4.2}
\end{center}
\end{figure}

Now we intend to give an interpretation of Fig.~\ref{fig:1.1} in view of the results
obtained in the previous section.
The dispersion of the points plotted in Fig.~\ref{fig:1.1} can be traced back to the spurious effect
introduced by the fact that many different spreads $s$ give their contribution to the final average. 
Therefore the next step is to investigate the frequency of odd spreads conditioned to a given $s$, in order to 
compare Eq.~\ref{eq:3.1} to real data.
\\
We can identify the variations of the spread caused by limit orders imposing the condition $s>s'$
because only limit orders inside the spread can decrease its value (the same argument has been also followed, for example, by \cite{lillo_spie}).
\begin{figure}[t]
\begin{center}
\includegraphics[clip=true,keepaspectratio,height=6.0cm,angle=0]{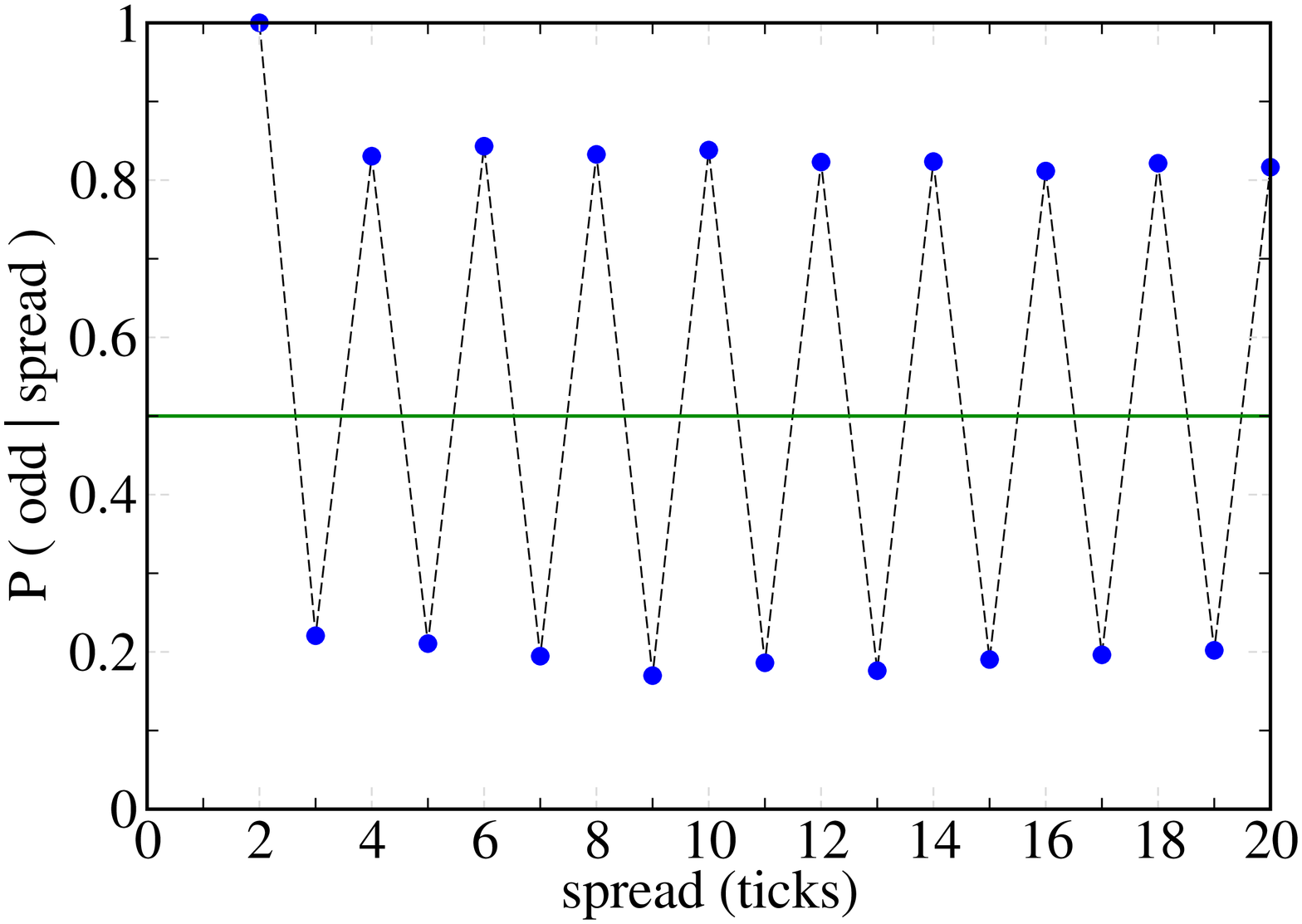}
\caption{(Color online) Conditional probability to have an odd-valued spread $s'$ given an initial
spread $s$ for real data. The plot shows clearly that with high probability an odd spread is followed 
by an even one ($0.8$) and vice-versa.}
\label{fig:4.1}
\end{center}
\end{figure}
In Fig.~\ref{fig:4.1} we show  the conditional probability to have an odd spread $s'$ starting from
a spread $s$, as a function of $s$. The pattern strongly oscillates around the value $1/2$: an even spread is most likely followed by an odd one as predicted by Eq.~\ref{eq:3.1}, but surprisingly also the viceversa is true.
\\
This apparently strange (with respect to the result of Eq. \ref{ashgdjs}) behavior can be attributed to a non-uniform depositions of limit orders inside the spread. 
We studied the distribution of the spread variations conditioned to a given value
of the spread $s$ for real data. We found that a consistent fraction $\alpha \sim 0.7$ of the limit orders inside the spread is placed at the quote adjacent to the best, as we show in Fig.~\ref{fig:4.2}.
A reasonable way to model this non-uniform distribution of the limit orders inside the spread
is through a piecewise constant function. 
In this way the probability to put a limit order at the first adjacent quote is $\alpha$, and the probability to
put the order in one of the other remaining quotes is equally distributed.
This tendency can be also interpreted in terms of agents' strategies. 
In fact, the placement an order far from the (previous) best bid or ask can be seen as a risky operation in which the agent, by disagreeing with other agents' evaluations, tries to trade quickly paying a kind of virtual cost equal to the distance between the best and her order quote. In other words, an order near the best is the most conservative position able to change the spread.
\begin{figure}[t]
\begin{center}
\includegraphics[clip=true,keepaspectratio,height=6.0cm,angle=0]{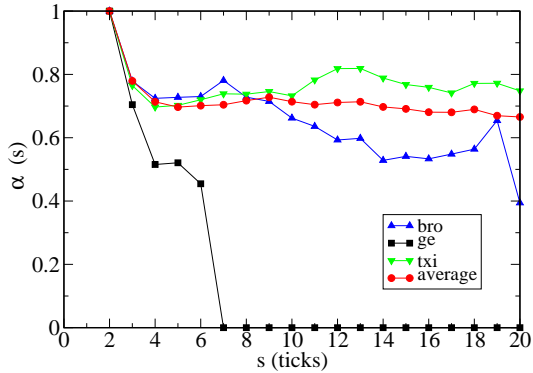}
\caption{(Color online) Probability $\alpha(s)$ to place an order at the quote adjacent to the best one
as a function of the spread. We have plotted this probabilities for three representative
stocks and also an average over all the 20 stocks. The very different values observed can be 
explained by considering the different statistics of the stocks. To properly address this effect we have performed an average over the 20 stocks of our data set. The result is a value of $\alpha(s)$
approximately constant.}
\label{fig:4.3}
\end{center}
\end{figure}
The next step is to find how much the probability $\alpha$ to place a limit order near the corresponding best depends on the spread $s$ and on the stock considered. In Fig.~\ref{fig:4.3} we plot $\alpha$ as a function of $s$ for three stocks
which cover a wide range of capitalization, finding very different behaviors depending on the stocks and on the values of the spreads. One possible explaination for this variety of behaviors is the lack of statistics: in fact, liquid stocks (that is, stocks characterized by fast order execution and small transaction impact on the price) usually have small spreads, and hence the statistics for large spreads is poor, and viceversa for illiquid stocks.
We can appropriately weight the contribution of different stocks by averaging on all the data.
The resulting curve is approximately a constant as a function of the spread. 

\section{A model for limit order deposition:\\ non-uniform case}
\label{sec:5}

\begin{figure}[htbp]
\begin{center}
\includegraphics[clip=true,keepaspectratio,height=6.0cm,angle=0]{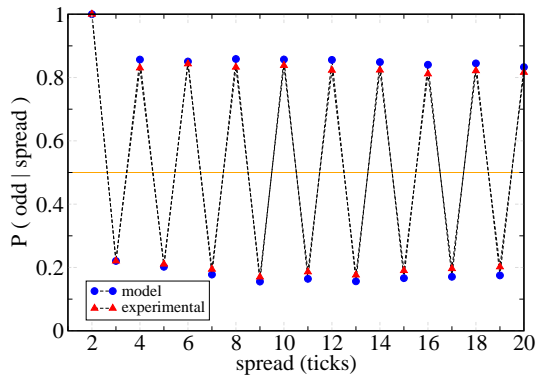}
\caption{(Color online) Comparison between the experimental data and the phenomenological model
described in the text. The phenomenological probabilities of the model (dots) show a 
small systematic overestimation of the oscillations with respect to the experimental
ones (triangles). This effect can be easily understood in terms of the approximation
considered in Eq.~\ref{eq:5.2}.}
\label{fig:5.1}
\end{center}
\end{figure}

\begin{figure}[htbp]
\begin{center}
\includegraphics[clip=true,keepaspectratio,height=6.0cm,angle=0]{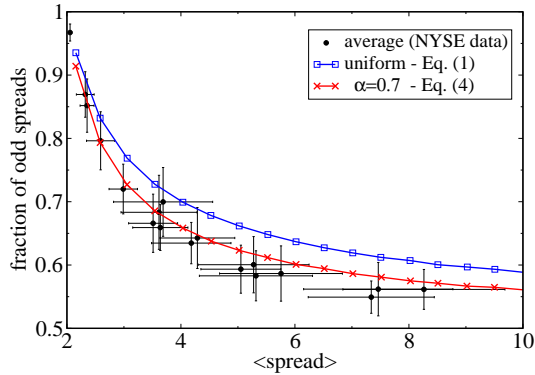}
\caption{(Color online) Fraction of odd-valued spread vs average spread for experimental data and for two different Monte 
Carlo simulations. The first simulation (squares) is performed using the uniform order deposition mechanism. The second one (crosses) is instead performed with the non uniform mechanism. These two simulations permit to investigate the two contributions to the asymmetry between odd and even spreads. As expected we observe that the
intrinsic asymmetry does not reproduce the experimental pattern (dots) while the two combined effects fit the experimental data very well. 
The small discrepancy found for large spreads is originated by the approximation made in Eq. \ref{eq:5.2}.}
\label{fig:5.4}
\end{center}
\end{figure}

In the previous section we have observed a systematic deviation of the experimental probabilities 
$P(o|e,s)$ and $P(e|o,s)$ from the ones of Eq.~\ref{eq:3.1} derived from the hypothesis of uniform order deposition.
In this section we are going to show that
this discrepancy is mainly due to the non-uniform 
probability of order placement inside the spread, and therefore to agents' strategies.\\
In order to include the effect of a non-uniform order deposition, we can generalize  Eq. \ref{eq:3.1} 
in the following way
\begin{equation}
\label{eq:5.1}
\left.
\begin{array}{l}
P(e|o,s)=\sum_{j=1}^{\frac{s-1}{2}}g(2j|s)\\\\
P(o|o,s)=\sum_{j=0}^{\frac{s-3}{2}}g(2j+1|s)
\end{array}
\right.
\left.
\begin{array}{l}
P(o|e,s)=\sum_{j=0}^{\frac{s-2}{2}}g(2j+1|s)\\\\
P(e|e,s)=\sum_{j=1}^{\frac{s-2}{2}}g(2j|s).
\end{array}
\right.
\end{equation}
where $g(i|s)$ is the probability mass function of the deposition mechanism for limit orders inside the spread and $i=1,\dots,s-1$ is the distance from the best quote of the placement price.
\\
In Fig.~\ref{fig:4.3} we have seen that the probability $\alpha$ that a limit order produces a spread variation equal to one
is weakly dependent on the value of the spread.
This suggests a simple approximation for $g(i|s)$ with a piecewise function. 
Now we discuss how to introduce this property in our model.
\\
If a limit order falls inside the book or at the best quotes the mechanism for order deposition is left unchanged.
Instead, if a limit order is placed inside the spread, the probabilities associated to the $s-1$ available quotes are no more uniform but highly peaked around the quote adjacent to the best. The deposition probabilities for a buy or a sell order, for a given $s$, become
\begin{equation}\label{eq:5.2}
g(i|s)=
\left\{
\begin{array}{l}
g(1|s)=\alpha \\\\
g(i|s)= \frac{1-\alpha}{s-2}\qquad i=2,...,s-1.
\end{array}
\right.
\end{equation}
For buy orders the index $i$ is the distance from the best bid while for sell orders the index $i$ is the distance from the best ask.
In the previous section we have seen that in real markets $\alpha$ can be approximately considered as constant and its value is about $0.7$. Clearly Eq. \ref{eq:5.2} is meaningful only for $s\geq3$ since for $s=1$ limit orders cannot fall inside the spread and for $s=2$ a limit order inside the spread is always placed at the quote adjacent to the best one.\\
It follows directly from Eq. \ref{eq:5.2} that the transition probabilities now read
\begin{equation}\label{eq:5.3}
\left.
\begin{array}{l}
P(e|o,s)=\alpha+\frac{1-\alpha}{2}\frac{s-3}{s-2}\\\\
P(o|o,s)=\frac{1-\alpha}{2}\frac{s-1}{s-2}
\end{array}
\right.
\left.
\begin{array}{l}
P(o|e,s)=\alpha+\frac{1-\alpha}{2}\\\\
P(e|e,s)=\frac{1-\alpha}{2}.
\end{array}
\right.
\end{equation}
\\
\newline
In Fig. \ref{fig:5.1} we plot a comparison between the expression of Eq. \ref{eq:5.3} and the corresponding probabilities evaluated from our data set. The experimental results are obtained by averaging over all the $80$ trading days and over all the $20$ stocks. 
The oscillating behavior can be explained by considering that a variation of spread of one tick ($\Delta s = 1$)
is highly preferred with respect to other variations. Hence an odd spread goes more likely to an even one
and vice-versa.
We are now able to separate the two effects that contribute to enhance the fraction of odd spreads and produce the pattern of Fig. \ref{fig:1.1} through a simple Monte Carlo simulation. These two contributions are the intrinsic asymmetry due to discreteness and non-uniformity of order deposition. As initial conditions we generate some sequences of spreads with different means, in order to represent different virtual stocks.
Starting from each sequence we simulate the transition to odd or even spreads $s'$ according to the probabilities of Eq. \ref{eq:3.1} and Eq. \ref{eq:5.3}. In such a way we can evaluate the average fraction of odd spreads for each virtual stock.
In Fig. \ref{fig:5.4} we compare the empirical average asymmetry with the results of the Monte Carlo simulations
in the two cases just mentioned.
The intrinsic asymmetry alone is not able to fit properly the experimental
data which instead are well reproduced by considering the two combined effects. 
Nevertheless we can observe some deviations for large spreads ($>6$).
This is due to the fact that we have assumed a constant probability to place order 
at quotes different from the one adjacent to the best.
When the spread grows the error introduced by this assumption becomes larger.
It is worth noticing that here we are neglecting the correlations between the spread values.
Anyway we can recover the experimental behavior even with this uncorrelated sequence since we are averaging on times far longer than the time scales of the spread correlation.

\section{The effects of the strategic deposition of orders}
\label{sec:6}

\begin{figure}[htbp]
\begin{center}
\includegraphics[clip=true,keepaspectratio,height=6.0cm,angle=0]{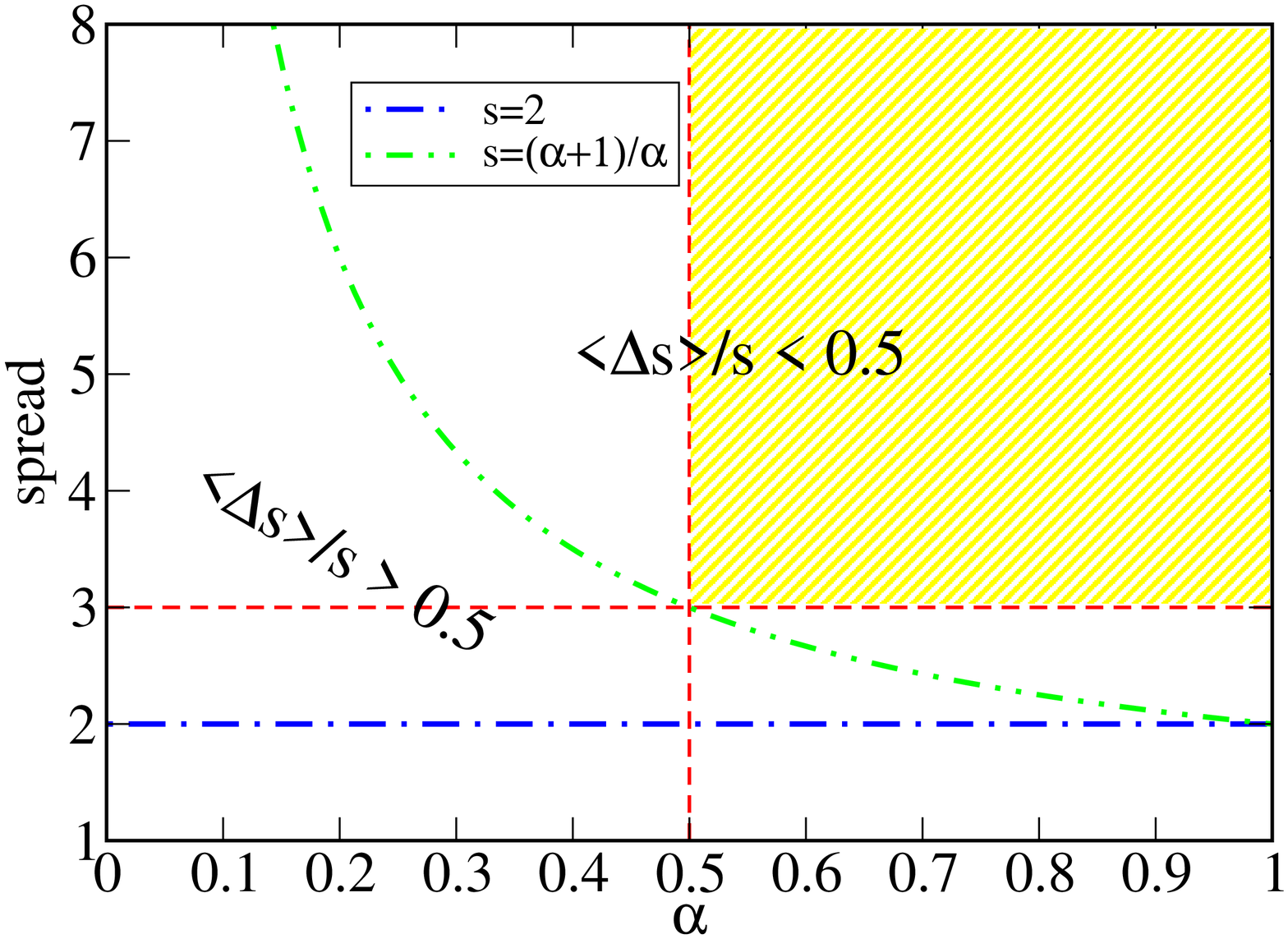}
\caption{Phase diagram of the relative average spread variation for the non-uniform mechanism.
The relative average spread variation $<\Delta s>/s$ is 0.5 for the uniform case.
$<\Delta s>/s$ is larger than $0.5$ only for spreads included in the region between $s=2$
and $s=(\alpha+1)/\alpha$. The highlighted region ($s\geq3$ and $\alpha>0.5$) corresponds
to a realistic scenario.
We see that in this region the relative average variation of the spread produced by a limit order is always smaller than 0.5. Therefore the non-uniform deposition reduces the coupling action of limit orders with respect to the uniform case.}
\label{fig:6.1}
\end{center}
\end{figure}

\begin{figure}[htbp]
\begin{center}
\includegraphics[scale=0.21]{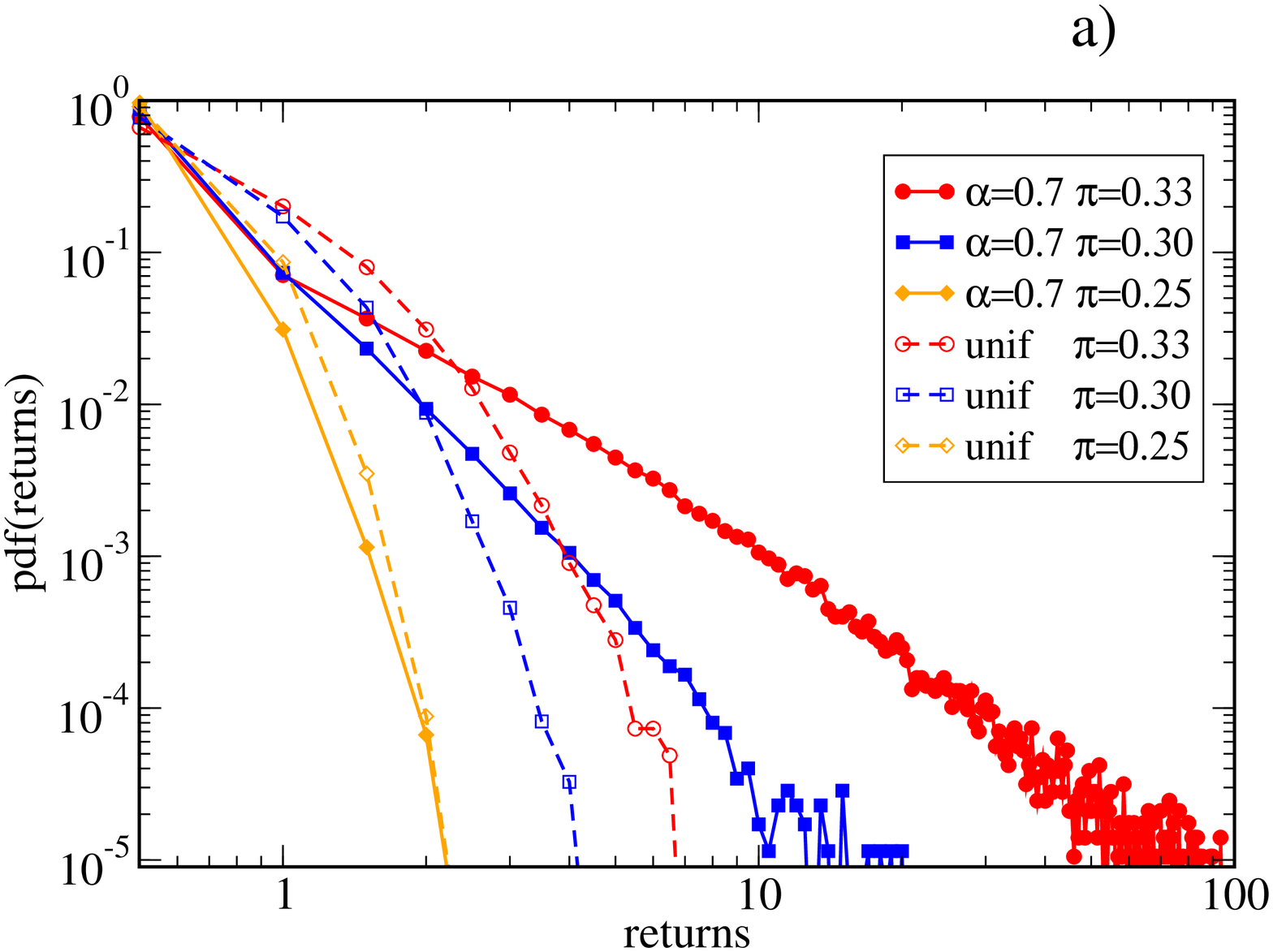}
\includegraphics[scale=0.21]{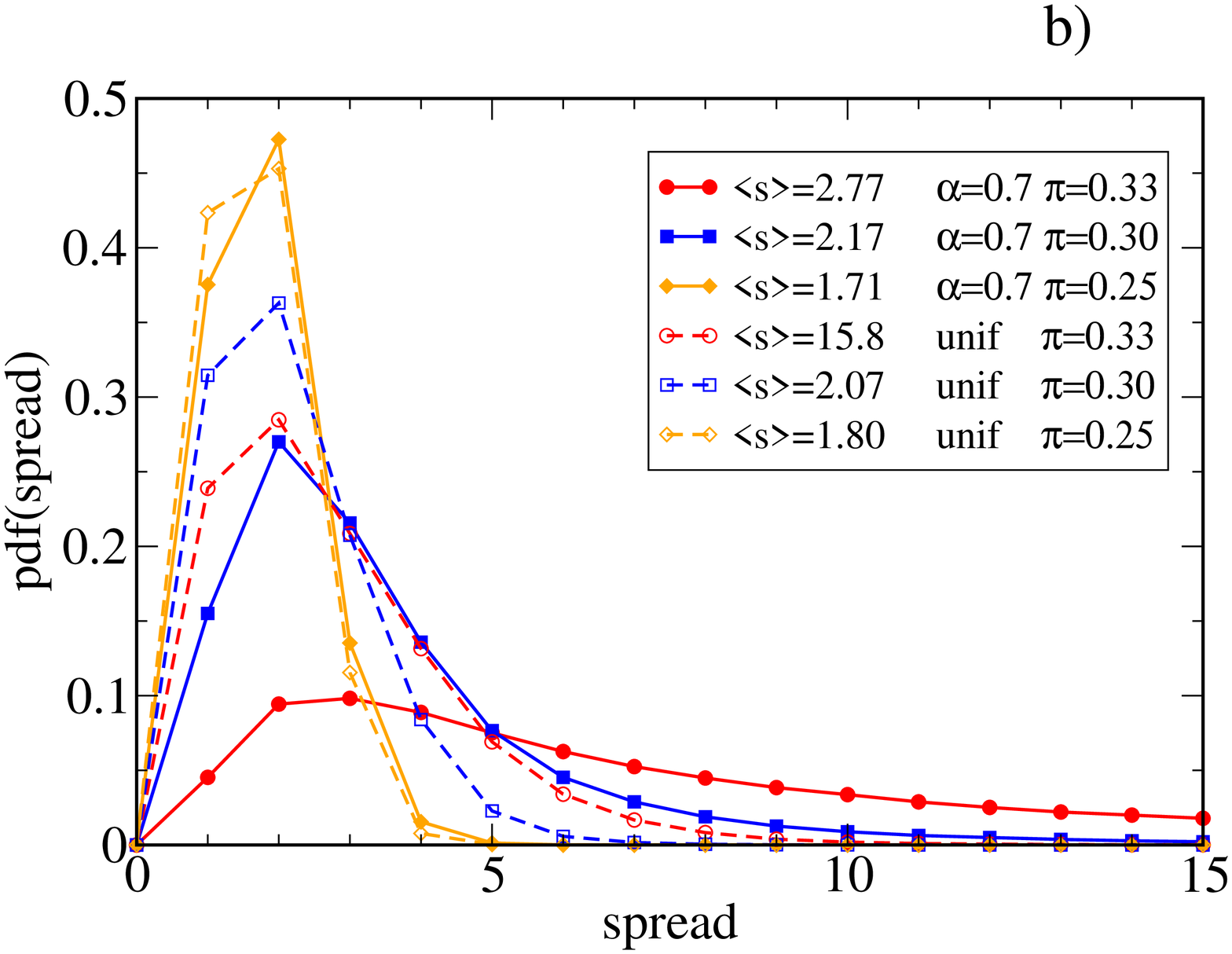}
\includegraphics[scale=0.21]{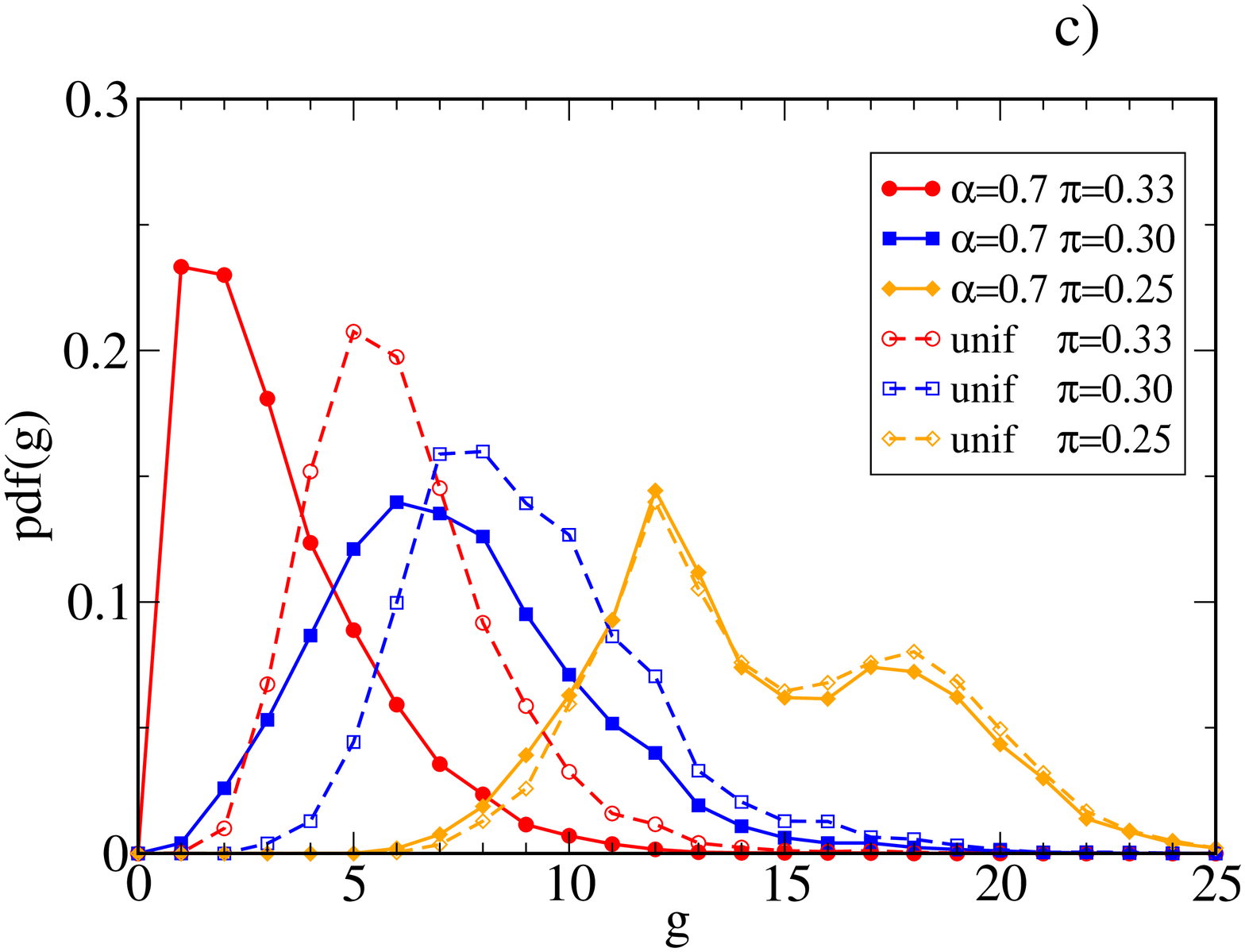}
\includegraphics[scale=0.21]{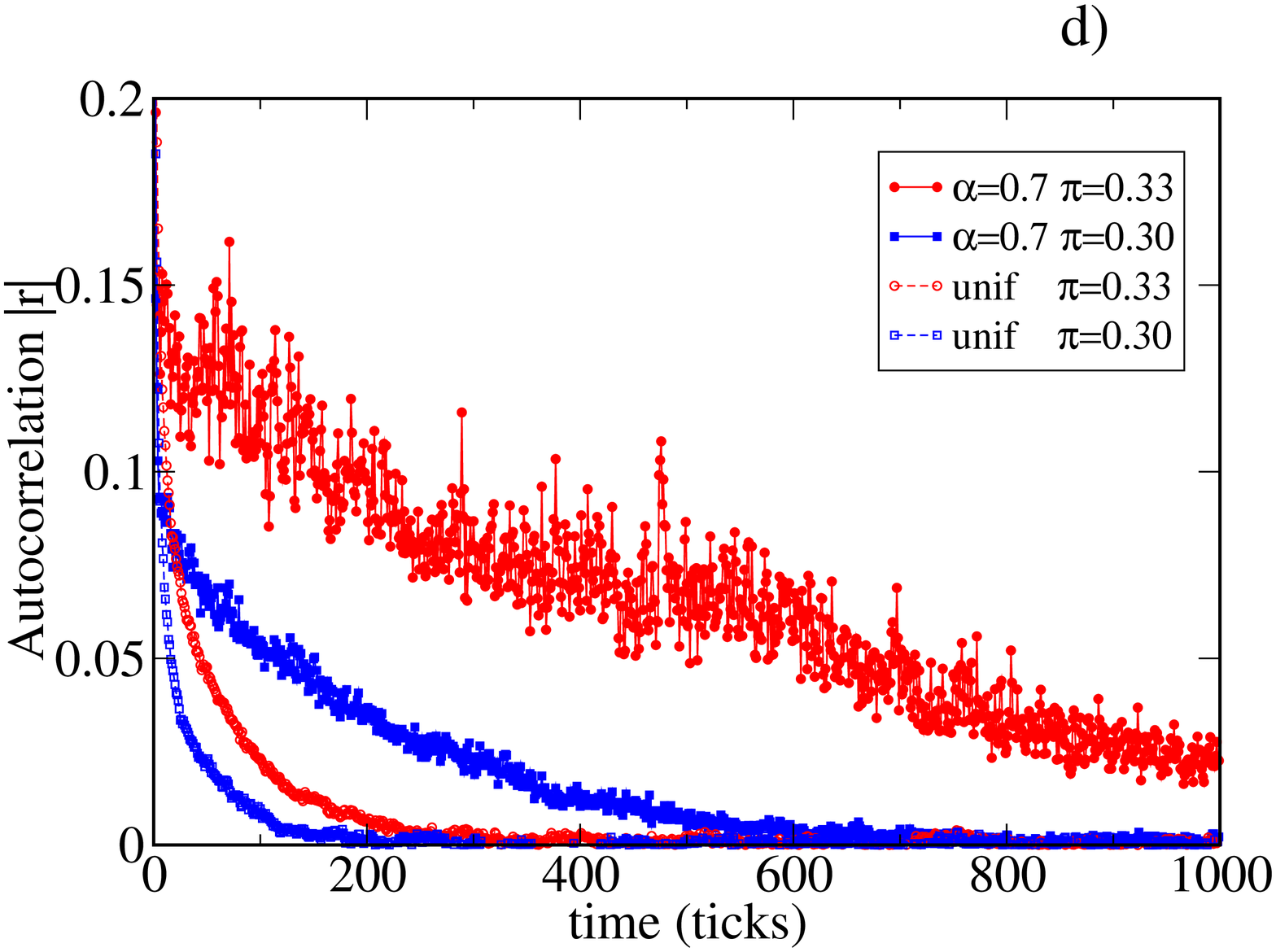}
\caption{(Color online) Statistical properties of the simulated order book for the uniform and non-uniform case. 
In panels a) and b) we plot the probability density functions
for the returns ($\Delta p$) and spreads respectively for different market order rates ($\pi$).
These plots show that the average fluctuations of spreads and returns are larger 
in the non-uniform case.
When the order book turns to a compact regime ($\pi\leq0.25$), the statistical properties of the model become nearly independent on the deposition details. Panel c) reveals that the non-uniform deposition produces non-trivial fluctuations of liquidity/granularity. 
In panel d) we plot the autocorrelation of the absolute values of returns and we observe the presence of persistent volatility. 
The decay of the autocorrelation function of the absolute values of returns is exponential except for short time lags where spurious effects take place.
This persistent behavior suggests the presence of an intermittent dynamics for the order book characterized by bursts of volatility.}
\label{fig:6.2}
\end{center}
\end{figure}

\begin{figure}[htbp]
\begin{center}
\includegraphics[clip=true,keepaspectratio,height=6.0cm,angle=0]{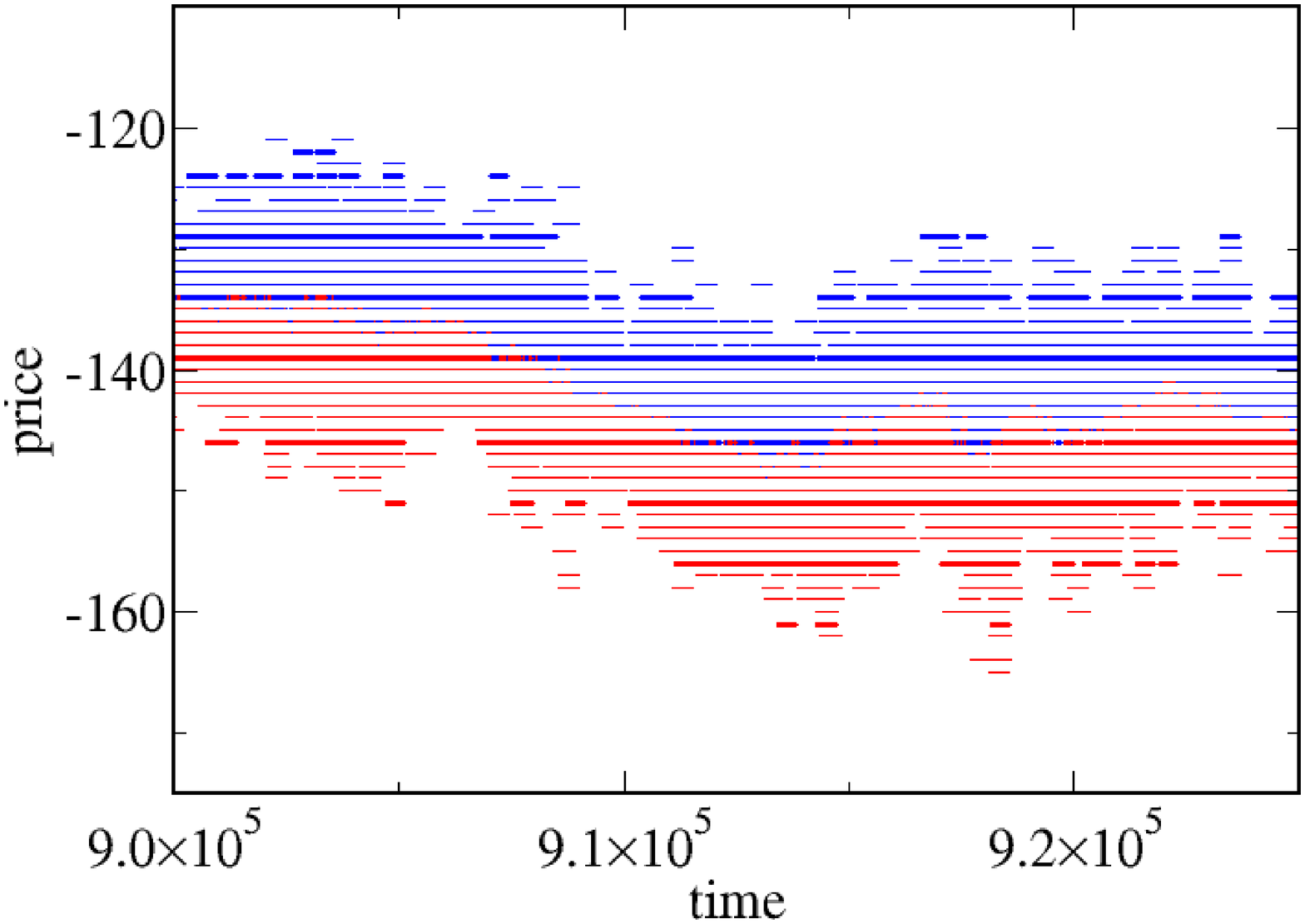}
\caption{(Color online) Snapshot of the simulated order book in the uniform case ($\pi=0.3$). The order book is always in a compact regime in which the average spread is nearly $1$ and a small and local volatility clustering is observed due to the {autoregressive} deposition rules. For higher values of $\pi$ small deviations from the compact regime appear but these phenomena cannot be compared to the intermittency produced when the order deposition
is non-uniform.}
\label{fig:6.3}
\end{center}
\end{figure}

\begin{figure}[htbp]
\begin{center}
\includegraphics[clip=true,keepaspectratio,height=6.0cm,angle=0]{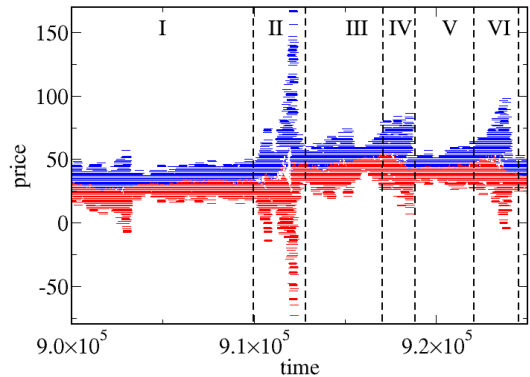}
\caption{(Color online) Snapshot of the simulated order book in the non-uniform case ($\pi=0.3$ and $\alpha=0.7$). These deposition rules produce an order book which is typically quiet and compact as in the uniform case (region I, III, V) but exhibits bursts of activity due to large fluctuations of the spread (region II, IV, VI). The system gives rise to a sort of intermittency since volatility is very persistent and clustered.}
\label{fig:6.4}
\end{center}
\end{figure}

An interesting question concerns the role of the parameter $\alpha$ and how the non-uniform deposition inside the spread affects the order book statistical properties. Our model allows this kind of investigation and permits to study the effect of different strategies of order placement inside the spread.\\ 
A numerical simulation reveals that if the same set of parameters of Sec. \ref{sec:3} and of~\cite{ob_pil} is used, the spread dynamics diverges for $\alpha>0.85$. 
To explain this effect we have to consider the interplay between market and limit orders. Market orders tend to move away the two best quotes eroding the book. Instead limit orders tend to reduce the spread by coupling the processes followed by best ask and best bid. In such a way the process for the spread is somehow stationary.
In this framework the deposition rules play an important role in softening or strengthening the coupling action of limit orders. In fact the coupling between $a(t)$ and $b(t)$ is ruled by two elements.
The first one is the fraction of orders which fall inside the spread. A larger fraction of these orders
produces a stronger coupling between the best quotes. 
The second is the mechanism of deposition of orders inside the spread, which governs the average spread variation produced by a limit order. 
In order to analyze the effect of the deposition mechanism we can reason as it follows.
By considering the symmetry of the uniform case, we find
\begin{equation}\label{eq:6.1}
\frac{<\Delta s>}{s}=\frac{1}{2}
\end{equation}
while in the non-uniform case, from Eq.~\ref{eq:5.3}, we obtain
\begin{equation}\label{eq:6.2}
\frac{<\Delta s>}{s}=\frac{\alpha}{s}+\frac{(1-\alpha)[s(s-1)-2]}{2s(s-2)}.
\end{equation}
We recall that $s\geq3$ since for $s=1$ limit orders cannot be placed inside the spread and for $s=2$ we always have $\Delta s/s=1/2$. 
The inequality $<\Delta s>/s \geq 1/2$ for Eq. \ref{eq:6.2} is satisfied when
\begin{equation}\label{eq:6.3}
2\leq s\leq \frac{\alpha+1}{\alpha}
\end{equation}
and we observe that $({\alpha+1})/{\alpha}>3$ only for $\alpha<0.5$ (see Fig. \ref{fig:6.1}). Consequently the average spread variation and the coupling action of limit orders in the non-uniform case are never larger than the one produced by the uniform mechanism in the region of parameters investigated ($s\geq3$ and $\alpha>0.5$).\\
Fixed $\alpha \leq0.8$ we can analyze the properties of our model and we will come back on the problem of the stability with respect to $\alpha$ at the end of this section. The set of figures in Fig. \ref{fig:6.2} clarifies the role of a non-uniform deposition inside the spread.
In Fig. \ref{fig:6.2}a and Fig. \ref{fig:6.2}b we have plotted the probability density functions
for the price variations (returns) and for the spreads respectively. 
We observe that in the non-uniform case the system produces larger fluctuations and larger
average spreads, as we expected from the previous discussion.
It is interesting to notice that when the system tends to a regime in which the order book is always compact, i.e. in which most of the quotes inside the book are occupied, the statistical properties becomes nearly independent on the deposition details ($\pi \leq 0.25$).
\\
Fig. \ref{fig:6.2}c reveals that the non-uniform deposition also amplifies the fluctuations of the granularity $g$, defined as the linear density of the volume stored in a side of the order book
~\cite{ob_pil}.
In particular the non-uniform deposition shows a non-trivial temporal structure for granularity that resembles an intermittency phenomenon. Since most of the arriving limit orders are placed adjacent to the best quote the order book stays for long times in a quiet and compact state characterized by an average spread whose value is nearly  one. This is the dominant regime of our simulated order book, but sometimes bursts of volatility are observed. In fact, when a large fluctuation of spread occurs, the {autoregressive} property and the non-uniformity of the limit order deposition make the relaxation towards the compact state very slow. This intermittency is directly related to the volatility correlation that is far longer in the non-uniform case than in the uniform one (see Fig. \ref{fig:6.2}d). We want to stress that a small volatility clustering is already present in the uniform case. Its origin can be traced back to the dependence on the past spread values of the deposition mechanism. This simple effect introduces an exponential correlation and so fixes a characteristic time-scale. The non-uniform deposition instead amplifies this effect because the mechanism sets a further and longer time-scale that depends on $\alpha$ (the correlation length increases for increasing values of $\alpha$).
The correlation functions in Fig. \ref{fig:6.2}d obey to an exponential decay except 
for short time lags where some spurious effects take place.
The bursts of volatility of the non-uniform case are even more evident if we represent the complete order book. 
In Fig. \ref{fig:6.3}, that corresponds to the uniform case, the order book is always compact. Instead in Fig. \ref{fig:6.4} we plot the non-uniform case and we find that the system stays for most of the time in a regime which is very similar 
to the one of Fig. \ref{fig:6.3} (regions I,III,V) but sometimes regions characterized by large spreads and large price movements appear (regions II,IV,VI).
It is worth noticing that in this model large price fluctuations emerge spontaneously, being triggered
by a random spread variation (and vice-versa). 
This mechanism resembles the phenomenon of self-organized criticality~\cite{bakbook,jensen}.
\\
It can be argued that also for larger values of $\pi$ the uniform case could produce this intermittency because this correspond to an increase of the time-scale on which the autoregressive mechanism is able to produce local volatility clustering as we can observe in Fig. \ref{fig:6.2}d. 
Nevertheless the correlation is still far shorter when $\pi=0.33$ with respect to the one generated by the non-uniform case with $\pi=0.3$. Furthermore the magnitude of the correlation is smaller 
than the one of the non-uniform case with lower values of $\pi$. Finally a visual inspection of the order book reveals that an intermittency of a kind appears but it is very small and we never obtain an order book with sudden transitions from a compact regime to a volatile regime as it happens in region II of Fig. \ref{fig:6.4}.\\
\\
Now we discuss the stability of the model for $\alpha \geq 0.8$.
We have seen in Figs. \ref{fig:6.2} a,b that, in order to make the system stable with respect to $\alpha$, a possible solution is the reduction of the probability of market orders $\pi$. In such a way it is possible to increase the average length of limit order sequences and then to
compensate the lower coupling. 
However increasing values of $\alpha$ ($>0.8$) would imply a choice for $\pi\approx 0.25$ (or even smaller) and in this region of parameters
the order book is always compact and all volatility bursts disappear. However it is worth noticing that these ranges of parameters are usually not observed in empirical data.

\section{Spread relaxation: role of the strategic order placement.}
\label{sec:7}
\begin{figure}[htbp]
\begin{center}
\includegraphics[clip=true,keepaspectratio,height=6.0cm,angle=0]{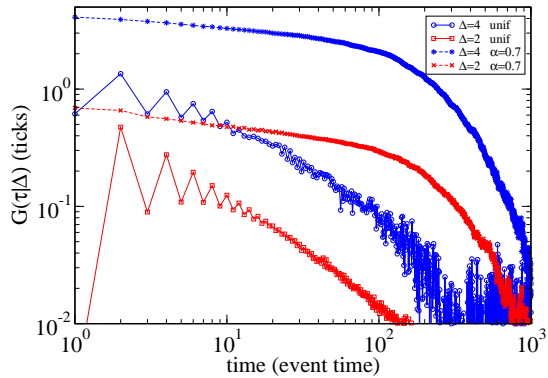}
\caption{(Color online) Spread decay given $\Delta$ in the uniform case (solid lines) and in the non-uniform case (dashed lines). We observe a much slower decay in the latter case. }
\label{fig:11}
\end{center}
\end{figure}

\begin{figure}[htbp]
\begin{center}
\includegraphics[clip=true,keepaspectratio,height=6.0cm,angle=0]{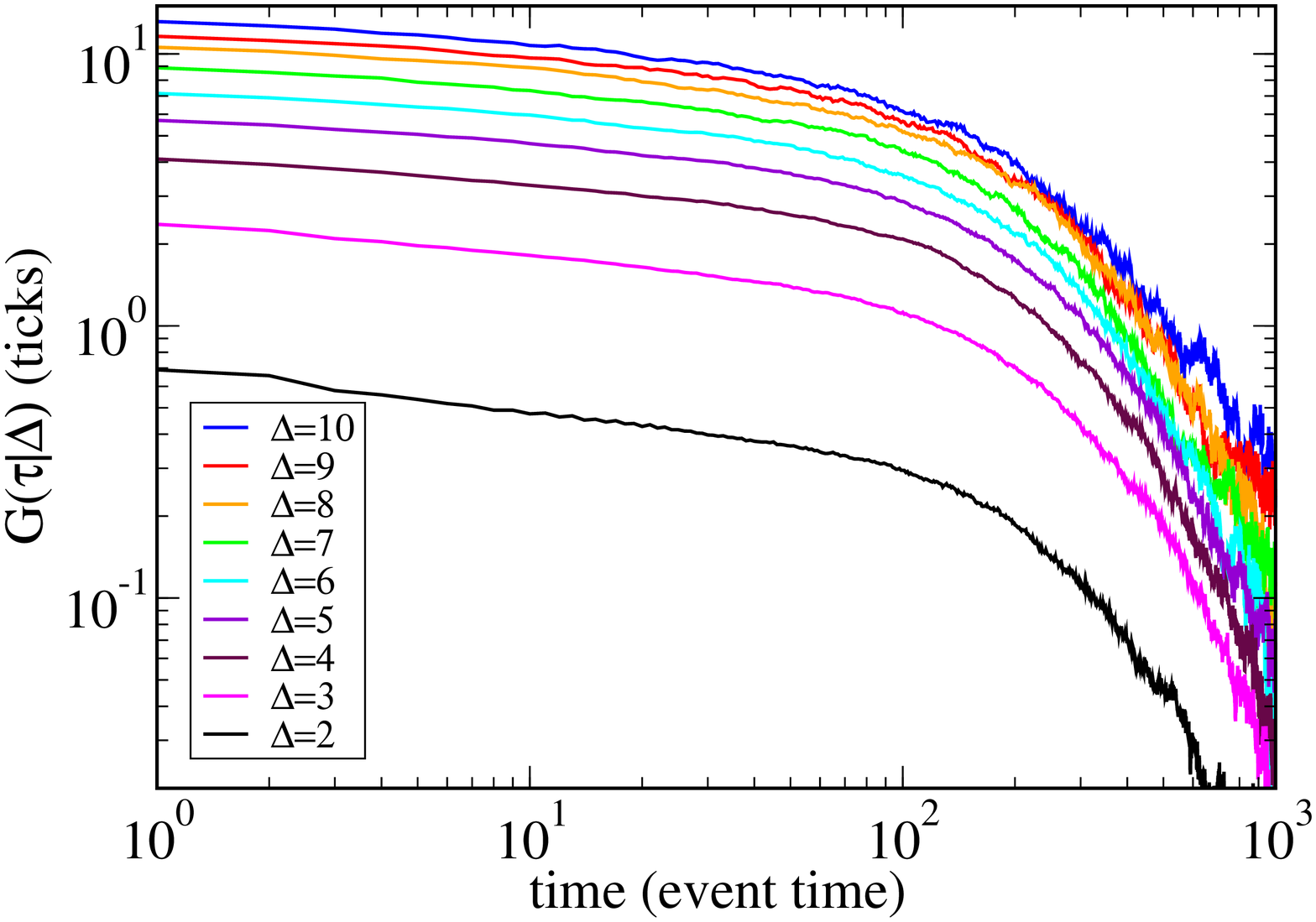}
\includegraphics[clip=true,keepaspectratio,height=6.0cm,angle=0]{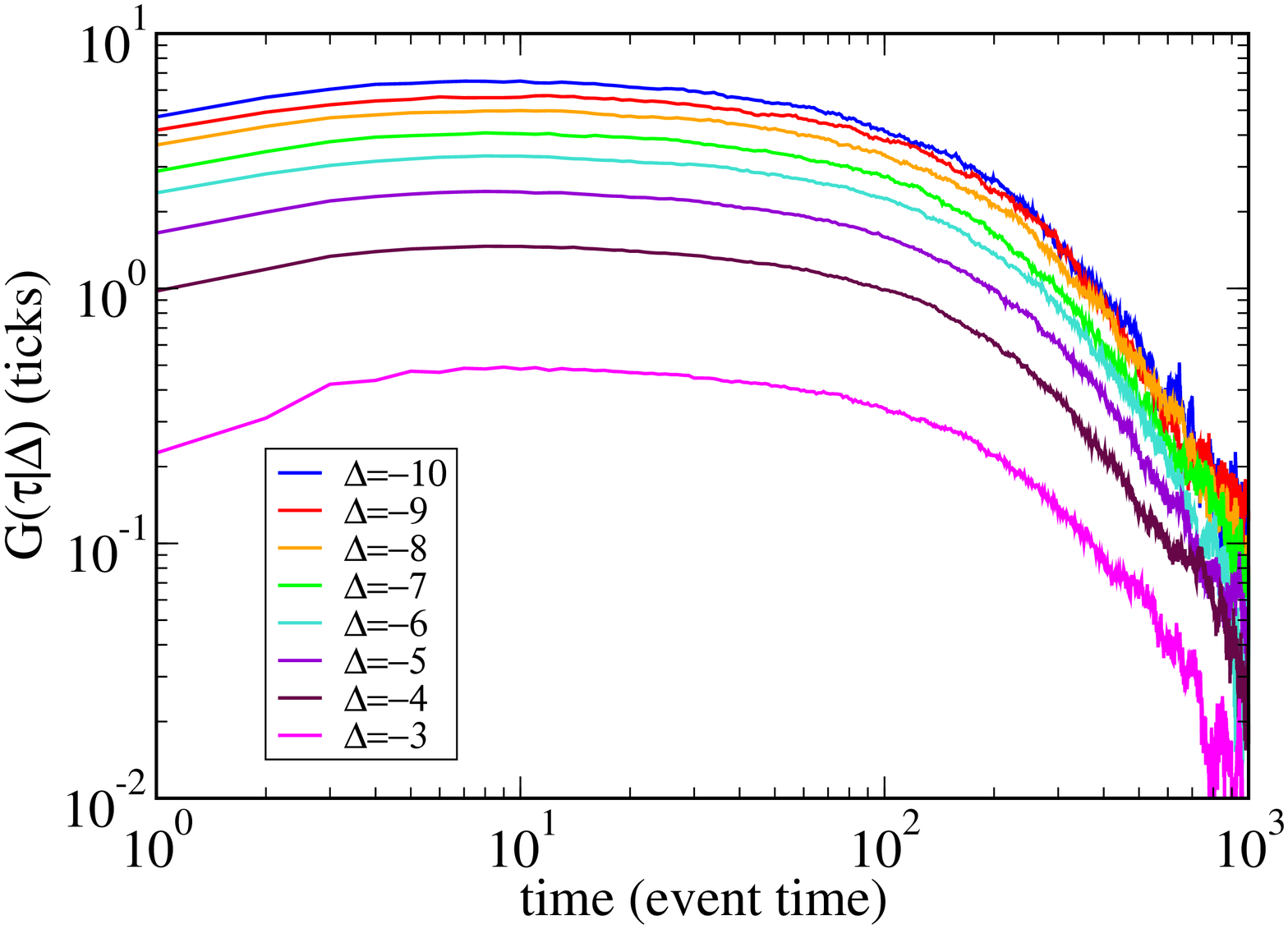}
\caption{(Color online) Spread decay in the non-uniform case for positive (top panel) and negative (bottom panel) values of $\Delta$. The value of $G(\tau|\Delta)$ is higher for higher absolute values of $\Delta$. The patterns found in our model are very similar to the empirical results of Ponzi et al. in \cite{ponzi}. }
\label{fig:12}
\end{center}
\end{figure}

Ponzi et al. in \cite{ponzi} studied the relaxation dynamics after an opening or a closing of the spread in LSE \footnote{London Stock Exchange} order book. They find a slow relaxation of the spread towards the mean value. This decay is compatible with a power law and the authors argue that the absence of a characteristic time scale is due to the presence of a strategic  placement of the orders. \\
We want to verify this empirical findings in the framework of the model introduced in the previous sections. In this respect we define, as in \cite{ponzi}, the quantity
\begin{equation}
G(\tau|\Delta) = E[s(t+\tau)|s(t)-s(t-1)=\Delta]-E[s(t)]
\end{equation}
where $E[\cdot]$ is the average on the whole time serie, $\Delta$ is the spread variation occured at time $t$ (i.e. $\tau=0$) and $\tau\geq0$. It is worth noticing that, in our model, $\tau$ is expressed in time event units differently from the analysis in \cite{ponzi} which is performed in physical time. 
The mapping between these two time units is not necessarily linear and therefore a quantitative agreement should not be always expected.\\
We perform the spread relaxation analysis in both the uniform case and non-uniform case and we report the results in Fig. \ref{fig:11}. 
The non-uniform mechanism produces a plateau of a kind and then for $\tau>100$ a faster decay to \textit{normal} spread values. Instead in the uniform case the spread relaxation is much faster than the previous case.
Now we analyze in detail the non-uniform case. In Fig. \ref{fig:12} we report the function $G(\tau|\Delta) $ for positive and negative values of $\Delta$ (corresponding respectively to openings and closings of the spread at time $\tau=0$). As in \cite{ponzi}, we observe two slightly different patterns for negative and positive $\Delta$. Instead, we do not observe such a difference in the uniform case, this means the nature of the relaxation dynamics is completely different for these two cases. \\
We can conclude that, in order to obtain a realistic spread relaxation function, a simple change to a pure zero-intelligence model consists in the non-uniform mechanism of limit orders deposition inside the spread described in section \ref{sec:5}. 

\section{Conclusions and perspectives}
\label{sec:con}

The order book is a system which is intrinsically discrete, for instance the quotes of placement of an order must be a multiple of the tick size.  
We have investigated which are the effects of this discreteness finding non-trivial aspects and deviations with respect to a continuous regime. 
\\
The starting point of this work  has been the observation that odd and even spreads are not equivalent for limit order deposition when the available quotes inside the spread are discrete. 
In fact, if a uniform deposition of orders inside the spread is assumed, the system spontaneously prefers odd 
spreads.
\\
One of results of this paper confirms that this asymmetry is present in real order books and that the fraction of odd spreads is significantly above $0.5$. 
However the asymmetry observed cannot be explained quantitatively only by considering the discrete nature of the order book.
Indeed we have found that a second effect also contributes to modulate the asymmetry,
the fact that agents prefer the quote adjacent to the best one when they place orders inside the spread.
\\
Both these contributions have been investigated in the framework of a microscopic model 
introduced in a previous work~\cite{ob_pil}.
The model permits to compare the effects of uniform and non-uniform deposition mechanisms for limit orders inside the spread. 
We have found that the asymmetry can be quantitatively reproduced in the framework of our model
by introducing a non-uniform deposition mechanism.\\
 Another result is the emergence of a sort of intermittent dynamics in which a regime characterized by a compact and liquid order book dominates but bursts of volatility also appear.
This intermittent behavior is also observed in real order books. In this respect, in the framework of our model, we compare the uniform and the non-uniform mechanism for order deposition with respect to the dynamics of the spread relaxation when a fluctuation occurs. We find that the introduction of a simple rule of order placement is sufficient to reproduce the peculiar pattern observed in real data (see \cite{ponzi}). The observed relaxation cannot be explained by a pure zero-intelligence model (see also \cite{toth}).
\\
An interesting point is how the agents' strategies take into account the discrete properties of the order book.
These question will be matter of future works.


\end{document}